\documentclass[twocolumn,superscriptaddress,nofootinbib,preprintnumbers,showpacs,tightenlines,notitlepage]{revtex4-1}
\usepackage[latin1]{inputenc}
\usepackage{blindtext}
\usepackage{graphicx}
\usepackage{amssymb}
\usepackage{color}
\usepackage{float}
\usepackage{amsmath}
\usepackage{amsfonts}
\usepackage{dcolumn}
\usepackage{hyperref}
\usepackage{amsthm}
\usepackage{color}
\usepackage{bm}

\def\uudot{\dot{u}}
\def\3nab{\tilde{\nabla}}

\def\la {\langle}
\def\ra {\rangle}

\def\be {\begin{equation}}
\def\ee {\end{equation}}
\def\ba {\begin{eqnarray}}
\def\ea {\end{eqnarray}}

\newtheorem*{thm*}{Theorem}
\newtheorem{theorem}{Theorem}

\newcommand{\bra}[1]{\left(#1\right)}

\newcommand{\sfr}[2]{{\textstyle\frac{#1}{#2}}}

\newcommand{\E}{{\mathcal E}}

\newcommand{\barray}{\begin{array}}
\newcommand{\earray}{\end{array}}
\newcommand{\e}{e}
\newcommand{\N}{N}

 \newcommand{\nab}{\nabla}

\newcommand{\del}{\nabla}

\newcommand \ep {\epsilon}

\newcommand{\udot}{{\mathcal A}}
\newcommand{\hh}{{\mathcal H}}

\def\ra{\rangle}
\def\la {\langle}

\newcommand{\bw}{\begin{widetext}}
\newcommand{\ew}{\end{widetext}}
\newcommand{\ben}{\begin{enumerate}}
\newcommand{\een}{\end{enumerate}}

\newcommand{\bef}{\begin{frame}}
\newcommand{\eef}{\end{frame}}
\newcommand{\bi}{\begin{itemize}}
\newcommand{\ei}{\end{itemize}}

\newcommand{\bfl}{\begin{flushleft}}
\newcommand{\efl}{\end{flushleft}}

\newcommand{\bb}{\begin{block}}
\newcommand{\eb}{\end{block}}

\newcommand{\bse}{\begin{subequation}}
\newcommand{\ese}{\end{subequation}}
\newcommand{\eei}{\end{eqnarray}\indent\indent}
\newcommand{\bc}{\begin{center}}
\newcommand{\ec}{\end{center}}
\newcommand{\ber}{\begin{eqnarray}}
\newcommand{\eer}{\end{eqnarray}}
\newcommand{\bern}{\begin{eqnarray*}}
\newcommand{\eern}{\end{eqnarray*}}
\newcommand{\beast}{\begin{equation*}}
\newcommand{\eeast}{\end{equation*}}
\newcommand{\bal}{\begin{align}}
\newcommand{\eal}{\end{align}} 
 
\newcommand \om {\omega}

\def\case#1/#2{\textstyle\frac{#1}{#2} }

\newcommand{\bver}{\begin{verbatim}}
\newcommand{\ever}{\end{verbatim}}

\begin{document}

\title{Conformal Symmetries of Locally Rotationally Symmetric Spacetimes}
\author{Sayuri Singh}
\email{sayurisingh22@gmail.com }
\affiliation{Astrophysics and Cosmology Research Unit, School of Mathematics, Statistics and Computer Science, University of KwaZulu-Natal, Private Bag X54001, Durban 4000, South Africa.}
\author{Rituparno Goswami}
\email{Goswami@ukzn.ac.za}
\affiliation{Astrophysics and Cosmology Research Unit, School of Mathematics, Statistics and Computer Science, University of KwaZulu-Natal, Private Bag X54001, Durban 4000, South Africa.}
\author{Sunil D. Maharaj}
\email{Maharaj@ukzn.ac.za}
\affiliation{Astrophysics and Cosmology Research Unit, School of Mathematics, Statistics and Computer Science, University of KwaZulu-Natal, Private Bag X54001, Durban 4000, South Africa.}

\begin{abstract}
In this paper we investigate conformal symmetries in Locally Rotationally Symmetric (LRS) spacetimes using a semitetrad covariant formalism. We demonstrate that a general LRS spacetime which rotates and spatially twists simultaneously has an inherent homothetic symmetry in the plane spanned by the fluid flow lines and the preferred spatial direction. We discuss the nature and consequence of this homothetic symmetry showing that a null Killing horizon arises when the heat flux has an extremal value. We also consider the special case of a perfect fluid and the restriction on the conformal geometry. 
\end{abstract}
 

\maketitle
%%%%%%%%%%%%%%%%%%%%%%%%%%%%%%%%%%%%%%%%%
\section{Introduction}
Locally rotationally symmetric spacetimes (LRS) are studied extensively in the literature because they contain metrics of physical interest  \cite{EllvanEll, Clarkson2003, Betschart2004, Clarkson:2007yp, Ellis_1968, Elst_Ellis_1996}. In particular, the spherically symmetric spacetimes are contained in this class. The LRS spacetimes have a preferred spatial direction at every point; a continuous isotropy group exists at each point which generates a multiply-transitive isometry group on the spacetime manifold. Consequently the semitetrad formalism, with a 1+1+2 decomposition, with covariantly defined scalar variables is an especially appropriate mathematical structure for describing the geometry and dynamics. Using the decomposition, Singh \textit{et al} \cite{Singh:2016qmr} found a new class of LRS spacetimes with nonvanishing rotation and spatial twist; this class has nonzero heat flux and is characterized with self-similarity. The most general rotating and twisting LRS spacetimes was found by Singh \textit{et al} \cite{Singh:2017qxi} which allows for a detailed analysis for gravitational collapse. Explicit forms of the LRS spacetime metrics and self-similar vectors were identified by Van den Bergh \cite{VandenBergh:2017}. \\

A conformal Klling vector has the property of preserving the spacetime metric up to a conformal factor. Conformal symmetries generate constants of the motion along null geodesics. General relativistic anisotropic fluids have been widely investigated in general relativity by Maartens \textit{et al} \cite{MaarMasonTsamp:1986}, Mason and Maartens \cite{MasonMaartens:1987} and Coley and Tupper \cite{ColeyTupper:1990} in the presence of a conformal symmetry. Conformal symmetries in static spherical spacetimes \cite{Maharajmaartens:1995, MaartensMahTupper:1995, MaartensMahTupper:1996, ManjMaharajMoop:2017, ManjMaharajMoop:2018}, shear-free spherical spacetimes \cite{MoopMaharaj:2013}, and general spherical spacetimes \cite{MoopMaharaj:2010} have been found and analysed. Spherically symmetric spacetimes are contained in the general LRS class and naturally the question of existence of conformal symmetries arises for this more general case. Particular studies have been undertaken in individual LRS Bianchi spacetimes, e.g. see the treatment of Khan \textit{et al} \cite{KhanHussainBok:2016}.  As far as we are aware, a general treatment of the conformal geometry in the broader LRS class has not been undertaken. Such an investigation in LRS spacetimesm using the 1+1+2 semitetrad formalism is likely to provide new insights into the spacetime geometry, the kinematics and the dynamics. This is the objective of our study. \\

The outline of the paper is summarized. In section 2 we discuss LRS spacetimes and outline the semitetrad formalism. The evolution, propagation and constraint equations are listed explicitly. We investigate the existence of a conformal symmetry in section 3. An explicit form of the conformal factor is found. This enables us to determine the nature of the conformal symmetry with rotation and spatial twist. The special case of perfect fluid spacetimes is considered in section 4 . The form of the conformal vector is constrained for a perfect fluid. Concluding remarks are made in section 5.

\section{Locally Rotationally Symmetric spacetimes in semitetrad formalism}
As discussed in the Introduction, Locally Rotationally Symmetric (LRS) spacetimes have a preferred spatial direction at every point in space. As described in detail in \cite{Ellis_1968}, if a spacetime exhibit local rotational symmetry in an open neighbourhood of a point $P$, then the coordinate freedoms can be used to describe the local metric in the neighbourhood in $(t,r,x,y)$ coordinates in the following way:
\ba\label{metric}
ds^2&=&-F^2(t,r)dt^2+X^2(t,r)dr^2\nonumber\\
&&+Y^2(t,r)[dx^2+D(x)dy^2]\nonumber\\
&&+g(x)F^2(t,r)[2dt-g(x)dy]dy\nonumber\\
&&-h(x)X^2(t,r)[2dr-h(x)dy]dy
\ea
Hence the $1+1+2$ semitetrad formalism \cite{EllvanEll,Clarkson2003,Betschart2004,Clarkson:2007yp} is very well suited to describe these spacetimes. In this formalism we decompose the spacetime covariantly in the following way
\be
g_{ab}=-u_au_b + h_{ab}=-u_au_b +e_ae_b +N_{ab}\;.
\ee
Here $u^a$ is the unit timelike vector along the matter flow lines with $u^a u_a = -1$ and $e^a$ is a spatial vector along the preferred direction with $e^a e_a = 1, u^a e_a = 0$.
The vector $u^{a}$ defines the \textit{covariant time derivative} along the flow lines (denoted by a dot) for any tensor 
$ S^{a..b}{}_{c..d}$ and is given by 
\be
\dot{S}^{a..b}{}_{c..d}{} = u^{e} \nab_{e} {S}^{a..b}{}_{c..d}\;.
\ee
The tensor $h_{ab}$ defines the fully orthogonally \textit{projected covariant derivative} $D$ for any tensor $
S^{a..b}{}_{c..d} $: 
\be D_{e}S^{a..b}{}_{c..d}{} = h^a{}_f
h^p{}_c...h^b{}_g h^q{}_d h^r{}_e \nab_{r} {S}^{f..g}{}_{p..q}\;,
\ee 
with total projection on all free indices. 
Therefore, the covariant derivative of $u^a$ is decomposed as
\be
\nabla_a u_b = -u_a A_b +\frac13\Theta h_{ab}+\sigma_{ab}+\ep_{abc}\om^{c},
\ee
where $A_b=\dot u_b$ is the acceleration, $\Theta=D_au^a$ is the expansion of $u_a$, $\sigma_{ab}=\bra{h^c{}_{\left( a \right.}h^d{}_{\left. b \right)}-\sfr13 h_{ab} h^{cd}}D_cu_d$ is the shear tensor (i.e. the rate of distortion) and $\om^{c}$ is the vorticity vector (i.e. the rotation). 
The Weyl tensor is split relative to $u^a$ into the \textit{electric} and \textit{magnetic Weyl curvature} parts as
\begin{eqnarray}
E_{ab} &=& C_{abcd}u^bu^d =E_{\la ab\ra}\;,
 \end{eqnarray}
and
\begin{eqnarray}
H_{ab} &=& \sfr12\ep_{ade}C^{de}{}_{bc}u^c =H_{\la ab\ra}  \;.
\end{eqnarray}
Similarly, the energy momentum tensor of matter is decomposed as 
\be
T_{ab}=\mu u_au_b+q_au_b+q_bu_a+ph_{ab}+\pi_{ab}\;,
\ee
where $\mu=T_{ab}u^au^b$ is the energy density, $q_a=q_{\la a\ra}=-h^{c}{}_aT_{cd}u^d$ is the 3-vector defining the heat flux, $p=(1/3 )h^{ab}T_{ab}$ is the isotropic pressure,   and $\pi_{ab}=\pi_{\la ab\ra}$ is the anisotropic stress. 

Further to this we can do a subsequent decomposition of the kinematical and dynamical quantities using the vector $e^a$.
This spacelike vector introduces two new derivatives, which for any tensor $ \psi_{a...b}{}^{c...d}  $: 
\ba
\hat{\psi}_{a..b}{}^{c..d} &\equiv & e^{f}D_{f}\psi_{a..b}{}^{c..d}~\label{eq:hat}, 
\\
\delta_f\psi_{a..b}{}^{c..d} &\equiv & N_{a}{}^{p}...N_{b}{}^gN_{h}{}^{c}..
N_{i}{}^{d}N_f{}^jD_j\psi_{p..g}{}^{i..j}\;.
\ea 
The hat-derivative (\ref{eq:hat}) is along the $e^a$ vector field in the surfaces orthogonal to $ u^{a}$ and the $\delta$-derivative (\ref{eq:delta}) is projected onto the 2-space perpendicular to both $u^a$ and $e^a$, which would be referred as ``sheet". This projection is orthogonal to $u^a$ and $e^a$ on every free index.
The 4-acceleration, vorticity and shear split as
\ba
\uudot^a&=&\udot \e^a+\udot^a,\\
\omega^a&=&\Omega \e^a+\Omega^a,\\
\sigma_{ab}&=&\Sigma\bra{\e_a\e_b-\sfr{1}{2}\N_{ab}}+2\Sigma_{(a}\e_{b)}+\Sigma_{ab},
\ea
and for the electric and magnetic Weyl tensors we have
\ba
E_{ab}&=&{\cal E}\bra{\e_a\e_b-\sfr{1}{2}\N_{ab}}+2{\cal E}_{(a}\e_{b)}+{\cal E}_{ab},\\
H_{ab}&=&{\cal H}\bra{\e_a\e_b-\sfr{1}{2}\N_{ab}}+2{\cal H}_{(a}\e_{b)}+{\cal
H}_{ab}.
\ea
The fluid variables $q^a$ and $\pi_{ab}$ are split as follows
\ba
q^a&=&Q \e^a+Q^a,\\
\pi_{ab}&=&\Pi\bra{\e_a\e_b-\sfr{1}{2}\N_{ab}}+2\Pi_{(a}\e_{b)}+\Pi_{ab}.
\ea
The covariant derivative of $e^a$ is decomposed in the directions orthogonal to $u^a$ into it's irreducible parts, which gives
\be 
 {D}_{a}e_{b} = e_{a}a_{b} + \frac{1}{2}\phi N_{ab} + 
\xi\epsilon_{ab} + \zeta_{ab}~.
\ee
Here, $\epsilon_{ab}=\epsilon_{[ab]}$ is the volume element on the sheet, $\phi$ is the \textit{spatial expansion of the sheet},  $\zeta_{ab}$ the \textit{spatial shear}, i.e., the distortion of the sheet, $a^{a}$ its \textit{spatial acceleration} , i.e., deviation from a geodesic, and $\xi$ is its spatial \textit{vorticity}, i.e., the ``twisting'' or rotation of the sheet. 

\subsection{Description of LRS spacetimes}

Because of the symmetries of LRS spacetimes, all the sheet vectors and tensors vanish identically:
\ba\label{eq:LRS}
\udot^a=\Omega^a =\Sigma_a={\cal E}_a={\cal H}_a=Q^a=\Pi^a=a_a=0,\\  \Sigma_{ab}={\cal E}_{ab}={\cal H}_{ab}=\Pi_{ab}=\zeta_{ab}=0.
\ea 
Therefore the remaining variables are
\ba\label{eq:LRSvar}
{\cal D}_1:&=&\{\udot,\Theta, \Omega, \Sigma, {\cal E}, {\cal H}, \mu, p, Q, \Pi,  \phi, \xi\}\\
&=& {\cal D}_{matter} + {\cal D}_{geometry},
\ea
where 
\be
{\cal D}_{matter}:=\{\mu, p, Q, \Pi\}\,,
\ee
are the matter variables that specify the energy momentum tensor of the matter, and 
\be
{\cal D}_{geometry}:=\{\udot,\Theta, \Omega, \Sigma, {\cal E}, {\cal H}, \phi, \xi\},
\ee
are the geometrical variables.
By decomposing the Ricci identities for $u^a$ and $e^a$ and the doubly contracted Bianchi identities, we get the field equations. They have the form \\
\medskip \\
\textit{Evolution}:
\ba
   \dot\phi &=& \bra{\sfr23\Theta-\Sigma}\bra{\udot-\sfr12\phi}
+2\xi\Omega+Q\ , \label{phidot}
\\ 
\dot\xi &=& \bra{\sfr12\Sigma-\sfr13\Theta}\xi+\bra{\udot-\sfr12\phi}\Omega
\nonumber \\ && +\sfr12 \hh,  \label{xidot}
\\
\dot\Omega &=& \udot\xi+\Omega\bra{\Sigma-\sfr23\Theta}, \label{dotomega}
\\
\dot \hh &=& -3\xi\E+\bra{\sfr32\Sigma-\Theta}\hh+\Omega Q
\nonumber\\ && +\sfr32\xi\Pi,
\ea
\smallskip

\textit{Propagation}:
\ba
\hat\phi  &=&-\sfr12\phi^2+\bra{\sfr13\Theta+\Sigma}\bra{\sfr23\Theta-\Sigma}
    \nonumber\\&&+2\xi^2-\sfr23\bra{\mu+\Lambda}
    -\E -\sfr12\Pi,\,\label{hatphinl}
\\
\hat\xi &=&-\phi\xi+\bra{\sfr13\Theta+\Sigma}\Omega , \label{xihat}
\\
\hat\Sigma-\sfr23\hat\Theta&=&-\sfr32\phi\Sigma-2\xi\Omega-Q\
,\label{Sigthetahat}
 \\
  \hat\Omega &=& \bra{\udot-\phi}\Omega, \label{Omegahat}
\\
\hat\E-\sfr13\hat\mu+\sfr12\hat\Pi&=&
    -\sfr32\phi\bra{\E+\sfr12\Pi}+3\Omega\hh
 \nonumber\\&&   +\bra{\sfr12\Sigma-\sfr13\Theta}Q , \label{Ehatmupi}
\\
\hat \hh &=& -\bra{3\E+\mu+p-\sfr12\Pi}\Omega
\nonumber\\&&-\sfr32\phi \hh-Q\xi,
\ea
\smallskip

\textit{Propagation/evolution}:
\ba
   \hat\udot-\dot\Theta&=&-\bra{\udot+\phi}\udot+\sfr13\Theta^2
    +\sfr32\Sigma^2 \nonumber\\
    &&-2\Omega^2+\sfr12\bra{\mu+3p-2\Lambda}\ ,\label{Raychaudhuri}
\\
    \dot\mu+\hat Q&=&-\Theta\bra{\mu+p}-\bra{\phi+2\udot}Q \nonumber \\
&&- \sfr32\Sigma\Pi,\,
\\    \label{Qhat}
\dot Q+\hat
p+\hat\Pi&=&-\bra{\sfr32\phi+\udot}\Pi-\bra{\sfr43\Theta+\Sigma} Q\nonumber\\
    &&-\bra{\mu+p}\udot\ ,
\ea %\\
\ba
\dot\Sigma-\sfr23\hat\udot
&=&
\sfr13\bra{2\udot-\phi}\udot-\bra{\sfr23\Theta+\sfr12\Sigma}\Sigma\nonumber\\
        &&-\sfr23\Omega^2-\E+\sfr12\Pi\, ,\label{Sigthetadot}
\\  
\dot\E +\sfr12\dot\Pi +\sfr13\hat Q&=&
    +\bra{\sfr32\Sigma-\Theta}\E-\sfr12\bra{\mu+p}\Sigma \nonumber \\
  && -\sfr12\bra{\sfr13\Theta+\sfr12\Sigma}\Pi+3\xi\hh \nonumber\\
    &&+\sfr13\bra{\sfr12\phi-2\udot}Q,
\label{edot}
\ea

\textit{Constraint:}
\be
\hh = 3\xi\Sigma-\bra{2\udot-\phi}\Omega. \label{H}
\ee

\subsection{Classification of LRS spacetimes}

An interesting property of LRS spacetimes is that {\it all} scalars $\Psi$ obey the following consistency relation:
\be\label{scalarcons}
\forall \Psi, \,\,\, \dot\Psi\Omega = \hat\Psi \xi,
\ee
The above can be proven as follows. We know for LRS spacetimes that
\ba
\nabla_au_b &=& -\udot u_ae_b+(\sfr13\Theta+\Sigma)e_ae_b+(\sfr13\Theta-\sfr12\Sigma)N_{ab} \nonumber\\
&& +\Omega\epsilon_{ab},\label{nablaub}\\
\nabla_ae_b &=& -\udot u_au_b+(\sfr13\Theta+\Sigma)e_au_b+\sfr12\phi N_{ab} \nonumber \\
&&+\xi\epsilon_{ab}, \label{nablaeb}
\ea
This immediately gives
\be
\Omega = \sfr12\epsilon^{ab}\nabla_au_b, \quad \xi = \sfr12\epsilon^{ab}\nabla_ae_b. \label{omega&xi}
\ee
Now for any scalar $\Psi$ in LRS spacetimes (where by the symmetry all derivatives on the sheet vanish) we have
\be
\nabla_b\Psi = -\dot\Psi u_b+\hat\Psi e_b.
\ee
Differentiating once more we get
\ba
\nabla_a\nabla_b\Psi &=& -(\nabla_a\dot\Psi)u_b-\dot\Psi(\nabla_au_b)+(\nabla_a\hat\Psi)e_b \nonumber \\
&&+\hat\Psi(\nabla_ae_b). 
\ea
Contracting the above with $\epsilon^{ab}$ the LHS becomes zero, as $\nabla_a\nabla_b\Psi$ is symmetric in $a,b$. Using equations (\ref{omega&xi}) the RHS becomes 
$-\dot\Psi(2\Omega)+\hat\Psi(2\xi)$, which proves equation (\ref{scalarcons}). Now using this equation we can immediately derive the consistency conditions for a perfect fluid form of matter with $Q=\Pi=0$. As described in \cite{Elst_Ellis_1996}, the propagation equations  evolve consistently in time if and only if 
\begin{equation}
\Omega\xi=0.
\end{equation} 
The above relation then naturally divides perfect fluid LRS spacetimes into three distinct subclasses \cite{Ellis_1968,Elst_Ellis_1996}: 
\begin{enumerate}
\item LRS class I: ($\Omega\neq 0, \xi=0$) These are stationary inhomogeneous rotating solutions.
\item LRS class II: ($\xi=0=\Omega$) These are inhomogeneous orthogonal family of solutions that can be both static or dynamic. Spherically symmetric solutions are contained in this class. 
\item LRS class III ($\xi\neq 0, \Omega= 0$):These are homogeneous orthogonal models with a spatial twist.
\end{enumerate}
In a couple of  recent papers \cite{Singh:2016qmr,Singh:2017qxi} we established the existence, and found the necessary and sufficient conditions, for the  general class of solutions of Locally Rotationally Symmetric spacetimes that have nonvanishing rotation and spatial twist simultaneously: that is for this class of spacetimes we %{\it must } 
have by definition 
\be\label{omegaxi}
\Omega\xi\neq0. 
\ee
The necessary condition for a LRS spacetime to have nonzero rotation and spatial twist simultaneously is the presence of nonzero heat flux $Q$ which is bounded from both sides.  
Also the equation (\ref{scalarcons}) generates further constraints and hence the total set of  constraint equations are now  $\mathcal{C}\equiv\{\mathcal{C}_1,\mathcal{C}_2,\mathcal{C}_3,\mathcal{C}_4\}$, where  
\ba
\mathcal{C}_1:= \hh& = & 3\xi\Sigma-\bra{2\udot+\frac{\Omega}{\xi}\left(\Sigma-\sfr23\Theta\right)}\Omega\,,  \label{constraint0}\\
\mathcal{C}_2:= \phi&=& -\frac{\Omega}{\xi}\left(\Sigma-\sfr23\Theta\right) \,, \label{constraint1} \\
\mathcal{C}_3:=Q&=&-\sfr{\sfr{\Omega}{\xi}}{1+\bra{\sfr{\Omega}{\xi}}^2}\left(\mu+p+\Pi\right),\label{omegaxi2} \\ 
\mathcal{C}_4:=\E &=&  \frac{\Omega}{\xi} \udot \left(\Sigma-\frac23\Theta\right)-\Sigma^2+\frac13\Theta\Sigma +\frac29\Theta^2 \nonumber \\
&&+2\left(\xi^2-\Omega^2\right)+\sfr{\bra{\sfr{\Omega}{\xi}}^2}{1+\bra{\sfr{\Omega}{\xi}}^2}\left(\mu+p+\Pi\right) \nonumber \\
&&-\frac12\Pi-\frac23\mu\;. \label{E}
\ea
In the next section we will investigate in detail the nature of conformal symmetries in this general spacetime.

\section{Conformal Symmetries in LRS spacetimes}

In this section we a priori assume that the spacetime has in general nonzero rotation and spatial twist, that is the matter field is in general imperfect. We now look for the existence of a conformal Killing vector $\mathbb{X}$ (with components $X^a$), which by definition satisfies the following condition
\be
L_{\mathbb{X}}g_{ab} = 2\psi g_{ab},
\ee
where $\psi$ is any scalar function. This in turn implies
\be 
\del_b X_a + \del_a X_b = 2\psi g_{ab}. \label{2psi}
\ee

In accordance with the symmetries of LRS spacetimes, where all the vectors and tensors projected to the 2-sheets vanish, let us look for a non-trivial conformal killing vector $\mathbb{X}$, in the [$u,e$] plane in the following form
\be
X_a = \alpha u_a + \beta e_a\;,  \label{Xa}
\ee
where $\alpha$ and $\beta$ are scalar functions of the local coordinates of the [$u,e$] plane.
With this choice of vector $\mathbb{X}$, (\ref{2psi}) becomes
\ba
2\psi g_{ab} &=& (\del_b\alpha)u_a + \alpha (\del_bu_a) +(\del_b\beta)e_a + \beta (\del_be_a) \nonumber \\
&&+  (\del_a\alpha)u_b + \alpha (\del_au_b) +(\del_a\beta)e_b + \beta (\del_ae_b). \label{2ps}
\ea
Now using equations (\ref{nablaub}) and (\ref{nablaeb}), and contracting with $u^au^b$, $e^ae^b$, $u^{(a}e^{b)}$ and $N^{ab}$, we get the following equations:
\ba
\dot\alpha+\beta\udot &=& \psi, \label{a}\\
\alpha(\sfr13\Theta+\Sigma)+\hat\beta &=& \psi, \label{b} \\
-\hat\alpha+\alpha\udot+\dot\beta-\beta(\sfr13\Theta+\Sigma) &=& 0, \label{c} \\
2\alpha(\sfr13\Theta-\sfr12\Sigma)+\beta\phi &=& 2\psi. \label{d}
\ea
Solving these equations for the scalars $\alpha,\beta$ and $\psi$, would give the required conformal Killing vector $\mathbb{X}$. We note here that if we only consider the above equations, then there are four equations for three unknowns, and hence the system is overdetermined. Therefore we cannot say anything about the existence of the solutions. However, owing to the symmetries of LRS spacetimes we have the property (\ref{scalarcons}), which further gives 
\be
\hat\alpha=\dot\alpha\frac{\Omega}{\xi} \quad  \mbox{and} \quad \dot\beta=\hat\beta\frac{\xi}{\Omega}. 
\ee
Substituting these equations into (\ref{c}) gives
\be
-\dot\alpha\frac{\Omega}{\xi}+\hat\beta\frac{\xi}{\Omega}+\alpha\udot-\beta(\sfr13\Theta+\Sigma)=0. \label{e}
\ee
From (\ref{a}) we have $\dot\alpha=\psi-\beta\udot$ and from (\ref{b}) we have $\hat\beta=\psi-\alpha(\sfr13\Theta+\Sigma)$. Substituting these equations into equation (\ref{e}), and solving for $\psi$ yields 
\be
\psi = \frac{\alpha\xi^2(\sfr13\Theta+\Sigma)-\beta\udot\Omega^2-\alpha\udot\xi\Omega+\beta\xi\Omega(\sfr13\Theta+\Sigma)}{\xi^2-\Omega^2}, \label{psi1}
\ee
whereas from equation (\ref{d}) we have
\be
\psi =\beta\left[\left(-\frac{1}{2}\frac{\alpha}{\beta}\right)\left(\Sigma-\frac{2}{3}\Theta\right)+\frac{1}{2}\phi\right]. \label{psi2}
\ee
Thus, we see that the equation (\ref{scalarcons}) reduces the system of conformal Killing differential equations to a system of two algebraic equations,  (\ref{psi1}) and (\ref{psi2}), with three unknowns. This is an underdetermined system and a solution is guaranteed. Equating (\ref{psi1}) and (\ref{psi2}), and solving for $\alpha / \beta$, we get
\be
\frac{\alpha}{\beta} = \frac{\sfr12\phi(\xi^2-\Omega^2)+\udot\Omega^2-\xi\Omega(\sfr13\Theta+\Sigma)}{\sfr32\Sigma\xi^2-\udot\xi\Omega+\Omega^2(\sfr13\Theta-\sfr12\Sigma)} \;,\label{alphabeta}
\ee
which is a consistency condition. This completes the proof for the existence of a conformal Killing vector in the [$u,e$] plane for a general LRS spacetimes with $\Omega\xi\neq0$. We now substitute the above in the constraint (\ref{d}), to get 
\be\label{psi}
  \psi =\,{\frac { \left( 3\,\Omega\,\Sigma-2\,\Omega\,\Theta+3\,\phi\,\xi \right) \beta\, \left( 3\,A\Omega-3\,\Sigma\,\xi-\Theta\,\xi \right)}{3(6\,A\Omega\,\xi+3\,{\Omega}^{2}\Sigma-2\,{\Omega}^{2}\Theta-9\,\Sigma\,{\xi}^{2})}}.
\ee
\\ Since the system is underdetermined, we always have the freedom to choose $\beta$, keeping $\alpha / \beta$ fixed. This freedom shows the arbitrariness of the magnitude of the vector $\mathbb{X}$, with a {\it fixed} direction. Now choosing 
\be
\beta= K\frac{3(6\,A\Omega\,\xi+3\,{\Omega}^{2}\Sigma-2\,{\Omega}^{2}\Theta-9\,\Sigma\,{\xi}^{2})}{ \left( 3\,\Omega\,\Sigma-2\,\Omega\,\Theta+3\,\phi\,\xi \right)\left( 3\,A\Omega-3\,\Sigma\,\xi-\Theta\,\xi \right)}\;,
\ee
where $K={\rm{const.}}$, we get $\psi=K$. In other words, we have the freedom to choose the magnitude of the conformal Killing vector in such a way that it becomes a {\it homothetic Killing vector}. This completes the proof for the theorem stated below:
\begin{theorem}
For a locally rotationally symmetric spacetime with simultaneous rotation and spatial twist, a homothetic Killing vector exists in the plane spanned by the fluid flow lines and the preferred spatial direction. Hence these spacetimes are self similar.
\end{theorem}
 The above theorem ensures that the field equations for these spacetimes can be written in terms of the variable $z= \tau/\rho$, where $\tau$ and $\rho$ are the curve parameters of the integral curves of $u$ and $e$ respectively.
 
 \subsection{Nature of the homothetic Killing vector}
 
From equation (\ref{Xa}) we have $X_a = \alpha u_a + \beta e_a$, therefore if  $X^aX_a<0$, the homothetic vector is timelike and if $X^aX_a>0$ it is spacelike. Now calculating $(\frac{\alpha^2}{\beta^2}-1$) we get the homothetic Killing vector is timelike (spacelike) iff
\ba
&&-\left[(\udot-\sfr12\phi-\sfr12(\Sigma-\sfr23\Theta))\Omega-\sfr12\xi(\phi+3\Sigma)\right](\xi-\Omega) \nonumber \\
&& (\Omega-\xi)\left[(\udot-\sfr12\phi+\sfr12(\Sigma-\sfr23\Theta))\Omega+\sfr12\xi(\phi-3\Sigma)\right] \nonumber \\
&& < (>)0.
\ea
Hence the condition for the homothetic Killing vector being null, generating a Killing horizon, is given by
\be\label{null}
\frac{\Omega}{\xi} = \pm1,
\ee
or
\be
\frac{\Omega}{\xi} = \frac{\phi\pm3\Sigma}{[-(\Sigma-\sfr23\Theta)\pm(2\udot-\phi)]}.
\ee
Note that a null hypersurface is a Killing horizon if the norm of the Killing vector field vanishes.

\subsection{Killing horizon}

We now discuss the case of a non-trivial null Killing horizon in the [$u,e$] plane. Comparing equations (\ref{psi1}) and (\ref{constraint1}), we can easily see that if 
$\alpha=-2\Omega, \beta=2\xi$, then $\psi=0$. In other words, if a homothetic vector exists with
\be
\frac{\alpha}{\beta} = -\frac{\Omega}{\xi},
\ee
then this vector will be a Killing horizon. To check the existence, we substitute the above equation in (\ref{alphabeta}), and get the following constraint
\be\label{null2}
\frac{\Omega^2}{\xi^2} = 1  \quad \Rightarrow \frac{\Omega}{\xi} = \pm1,
\ee
which implies, by equation (\ref{null}), that the Killing vector is null. Also from (\ref{omegaxi2}), we see that 
\be\label{Q}
\frac{\Omega}{\xi} = \frac{-\bra{\mu+p+\Pi}\mp \sqrt{\bra{\mu+p+\Pi}^2-4Q^2}}{2Q}.
\ee
In other words, the situation described by (\ref{null2}) can only happen when the heat flux attains the extremal values
\be
Q=\pm\frac12\bra{\mu+p+\Pi}.
\ee
Hence we can now state the following theorem:
\begin{theorem}
For a locally rotationally symmetric spacetime with simultaneous rotation and spatial twist, whenever the magnitude of the rotation equals the magnitude of spatial twist, the homothetic vector in the plane spanned by the fluid flow lines and the preferred spatial direction, becomes null, creating a Killing horizon. This happens when the bounded and nonzero heat flux attains extremal values. 
\end{theorem}
 
\section {Special cases of perfect fluid spacetimes}

In this section we discuss special cases, when matter is in the form of a perfect fluid. As we have already indicated for this case $\Omega\xi=0$. We now describe the general properties of the conformal Killing vector for all three subclasses of perfect fluid LRS spacetimes.

\begin{enumerate}

\item \textbf{LRS I:} In this case we have $\Omega\neq0$ and $\xi=0$. Therefore from the equation (\ref{scalarcons}), we see that for all scalars $\Psi$, we have $\dot\Psi=0$. Since the dot derivative of all scalars vanish, we must have $\Theta=\Sigma=0$. Now from equation (\ref{alphabeta}), we get
\be
\frac{\alpha}{\beta}=\frac{-\sfr12\phi+\udot}{\sfr13\Theta-\sfr12\Sigma}\rightarrow\infty.
\ee
Also from the equation (\ref{psi2}) we get 
\be
\psi=0.
\ee
Therefore, we see that in this case the homothetic vector becomes a timelike Killing vector. Hence the spacetime is stationary.

\item \textbf{LRS III:}  In this case we have $\xi\neq0$, $\Omega=0$. Again from equation (\ref{scalarcons}), we see that for all scalars $\Psi$, we have $\hat\Psi=0$. Since the hat derivative of all scalars vanish, we must have $\phi=0$. From equation (\ref{alphabeta}), we get
\be
\frac{\alpha}{\beta}=\frac{1}{3}\frac{\phi}{\Sigma}\rightarrow 0.
\ee
and 
\be
\psi=0.
\ee
Hence in this case the homothetic vector becomes a spacelike Killing vector. Consequently the spacetime is spatially homogeneous. 

\item\textbf{LRS II:} This is the subclass where both $\Omega=0$ and $\xi=0$. Spherically symmetric spacetimes are contained in this subclass. We can easily see that this subclass is a singular point in our analysis, and the homothetic vector is no longer well defined. This is due to the fact that LRS II spacetimes do not contain any inherent Killing symmetries in the [$u,e$] plane. Killing symmetries have to be imposed separately to find static, homogeneous, self similar or conformally symmetric classes of solutions. \\
\end{enumerate}

\section{Conclusion}

In this paper we have performed a detailed study of conformal symmetries in LRS spacetimes utilizing the semitetrad formalism. We proved that locally rotationally symmetric (LRS) spacetimes with simultaneous rotation and spatial twist, are always self similar as they have a homothetic Killing vector in the plane spanned by the fluid flow lines and the preferred spatial direction. This homothetic Killing vector becomes a null Killing horizon in the special case when the nonzero and bounded heat flux reaches an extremal value. Also when either the rotation or the spatial twist vanish, this homothetic vector becomes a timelike or spacelike Killing vector making the spacetime either stationary or spatially homogeneous. However when both these quantity vanish there is no inherent Killing or conformal symmetry in the plane spanned by the fluid flow lines and the preferred spatial direction.

\end{document}